\documentclass[twocolumn,superscriptaddress,showpacs,preprintnumbers,amsmath,amssymb,nofootinbib]{revtex4-1}

\usepackage{graphicx}
\usepackage{dcolumn}
\usepackage{bm}
\usepackage[dvips]{color}
\usepackage{epsfig}

\RequirePackage{xspace}

\newcommand{\dsdpi}{\ensuremath{D^{*+}\to D^0\pi^+}\xspace}

\newcommand{\kspp}{\ensuremath{K^0_S\pi^+\pi^-}\xspace}
\newcommand{\dkpp}{\ensuremath{D\to K^0_S\pi^+\pi^-}\xspace}
\newcommand{\dnkpp}{\ensuremath{D^0\to K^0_S\pi^+\pi^-}\xspace}
\newcommand{\dbkpp}{\ensuremath{\overline{D}{}^0\to K^0_S\pi^+\pi^-}\xspace}

\newcommand{\bdk}{\ensuremath{B^{\pm}\to D K^{\pm}}\xspace}

\newcommand{\bdkm}{\ensuremath{B^-\to D K^-}\xspace}
\newcommand{\bdkp}{\ensuremath{B^+\to D K^+}\xspace}

\newcommand{\bdpi}{\ensuremath{B^{\pm}\to D \pi^{\pm}}\xspace}

\newcommand{\bdpim}{\ensuremath{B^-\to D \pi^-}\xspace}
\newcommand{\bdpip}{\ensuremath{B^+\to D \pi^+}\xspace}

\newcommand{\dn}{\ensuremath{D^0}\xspace}
\newcommand{\dnbar}{\ensuremath{\overline{D}{}^0}\xspace}

\newcommand{\ab}{\ensuremath{A_B}\xspace}

\newcommand{\pd}{\ensuremath{P}\xspace}
\newcommand{\pdbar}{\ensuremath{\overline{P}}\xspace}

\newcommand{\ad}{\ensuremath{A}\xspace}
\newcommand{\adbar}{\ensuremath{\overline{A}}\xspace}
\newcommand{\dvar}{\ensuremath{(m^2_{+}, m^2_{-})}\xspace}

\newcommand{\aad}{\ensuremath{|A|}\xspace}
\newcommand{\aadbar}{\ensuremath{|\overline{A}|}\xspace}

\newcommand{\mbc}{\ensuremath{M_{\rm bc}}\xspace}
\newcommand{\de}{\ensuremath{\Delta E}\xspace}
\newcommand{\thr}{\ensuremath{\cos\theta_{\rm thr}}\xspace}
\newcommand{\fish}{\ensuremath{\mathcal{F}}\xspace}

\newcommand{\ddd}{\ensuremath{\Delta\delta_D}\xspace}

\begin{document}

\preprint{BELLE-CONF-1103}

\title{First Measurement of \boldmath{$\phi_3$} with a Binned 
Model-independent Dalitz Plot Analysis 
of \boldmath{\bdk}, \boldmath{\dkpp} Decay}

\affiliation{University of Bonn, Bonn}
\affiliation{Budker Institute of Nuclear Physics, Novosibirsk}
\affiliation{Faculty of Mathematics and Physics, Charles University, Prague}
\affiliation{Chiba University, Chiba}
\affiliation{University of Cincinnati, Cincinnati, Ohio 45221}
\affiliation{Department of Physics, Fu Jen Catholic University, Taipei}
\affiliation{Justus-Liebig-Universit\"at Gie\ss{}en, Gie\ss{}en}
\affiliation{Gifu University, Gifu}
\affiliation{The Graduate University for Advanced Studies, Hayama}
\affiliation{Gyeongsang National University, Chinju}
\affiliation{Hanyang University, Seoul}
\affiliation{University of Hawaii, Honolulu, Hawaii 96822}
\affiliation{High Energy Accelerator Research Organization (KEK), Tsukuba}
\affiliation{Hiroshima Institute of Technology, Hiroshima}
\affiliation{University of Illinois at Urbana-Champaign, Urbana, Illinois 61801}
\affiliation{Indian Institute of Technology Guwahati, Guwahati}
\affiliation{Indian Institute of Technology Madras, Madras}
\affiliation{Indiana University, Bloomington, Indiana 47408}
\affiliation{Institute of High Energy Physics, Chinese Academy of Sciences, Beijing}
\affiliation{Institute of High Energy Physics, Vienna}
\affiliation{Institute of High Energy Physics, Protvino}
\affiliation{Institute of Mathematical Sciences, Chennai}
\affiliation{INFN - Sezione di Torino, Torino}
\affiliation{Institute for Theoretical and Experimental Physics, Moscow}
\affiliation{J. Stefan Institute, Ljubljana}
\affiliation{Kanagawa University, Yokohama}
\affiliation{Institut f\"ur Experimentelle Kernphysik, Karlsruher Institut f\"ur Technologie, Karlsruhe}
\affiliation{Korea Institute of Science and Technology Information, Daejeon}
\affiliation{Korea University, Seoul}
\affiliation{Kyoto University, Kyoto}
\affiliation{Kyungpook National University, Taegu}
\affiliation{\'Ecole Polytechnique F\'ed\'erale de Lausanne (EPFL), Lausanne}
\affiliation{Faculty of Mathematics and Physics, University of Ljubljana, Ljubljana}
\affiliation{Luther College, Decorah, Iowa 52101}
\affiliation{University of Maribor, Maribor}
\affiliation{Max-Planck-Institut f\"ur Physik, M\"unchen}
\affiliation{University of Melbourne, School of Physics, Victoria 3010}
\affiliation{Nagoya University, Nagoya}
\affiliation{Nara University of Education, Nara}
\affiliation{Nara Women's University, Nara}
\affiliation{National Central University, Chung-li}
\affiliation{National United University, Miao Li}
\affiliation{Department of Physics, National Taiwan University, Taipei}
\affiliation{H. Niewodniczanski Institute of Nuclear Physics, Krakow}
\affiliation{Nippon Dental University, Niigata}
\affiliation{Niigata University, Niigata}
\affiliation{University of Nova Gorica, Nova Gorica}
\affiliation{Novosibirsk State University, Novosibirsk}
\affiliation{Osaka City University, Osaka}
\affiliation{Osaka University, Osaka}
\affiliation{Pacific Northwest National Laboratory, Richland, Washington 99352}
\affiliation{Panjab University, Chandigarh}
\affiliation{Peking University, Beijing}
\affiliation{Princeton University, Princeton, New Jersey 08544}
\affiliation{Research Center for Nuclear Physics, Osaka}
\affiliation{RIKEN BNL Research Center, Upton, New York 11973}
\affiliation{Saga University, Saga}
\affiliation{University of Science and Technology of China, Hefei}
\affiliation{Seoul National University, Seoul}
\affiliation{Shinshu University, Nagano}
\affiliation{Sungkyunkwan University, Suwon}
\affiliation{School of Physics, University of Sydney, NSW 2006}
\affiliation{Tata Institute of Fundamental Research, Mumbai}
\affiliation{Excellence Cluster Universe, Technische Universit\"at M\"unchen, Garching}
\affiliation{Toho University, Funabashi}
\affiliation{Tohoku Gakuin University, Tagajo}
\affiliation{Tohoku University, Sendai}
\affiliation{Department of Physics, University of Tokyo, Tokyo}
\affiliation{Tokyo Institute of Technology, Tokyo}
\affiliation{Tokyo Metropolitan University, Tokyo}
\affiliation{Tokyo University of Agriculture and Technology, Tokyo}
\affiliation{Toyama National College of Maritime Technology, Toyama}
\affiliation{CNP, Virginia Polytechnic Institute and State University, Blacksburg, Virginia 24061}
\affiliation{Wayne State University, Detroit, Michigan 48202}
\affiliation{Yamagata University, Yamagata}
\affiliation{Yonsei University, Seoul}
  \author{I.~Adachi}\affiliation{High Energy Accelerator Research Organization (KEK), Tsukuba} 
  \author{K.~Adamczyk}\affiliation{H. Niewodniczanski Institute of Nuclear Physics, Krakow} 
  \author{H.~Aihara}\affiliation{Department of Physics, University of Tokyo, Tokyo} 
  \author{K.~Arinstein}\affiliation{Budker Institute of Nuclear Physics, Novosibirsk}\affiliation{Novosibirsk State University, Novosibirsk} 
  \author{Y.~Arita}\affiliation{Nagoya University, Nagoya} 
  \author{D.~M.~Asner}\affiliation{Pacific Northwest National Laboratory, Richland, Washington 99352} 
  \author{T.~Aso}\affiliation{Toyama National College of Maritime Technology, Toyama} 
  \author{V.~Aulchenko}\affiliation{Budker Institute of Nuclear Physics, Novosibirsk}\affiliation{Novosibirsk State University, Novosibirsk} 
  \author{T.~Aushev}\affiliation{Institute for Theoretical and Experimental Physics, Moscow} 
  \author{T.~Aziz}\affiliation{Tata Institute of Fundamental Research, Mumbai} 
% ...
\collaboration{The Belle Collaboration}


\begin{abstract} 

We present the first measurement of the angle $\phi_3$ of the 
unitarity triangle using a binned model-independent Dalitz plot analysis 
technique of \bdk, \dkpp decay chain. The method is based on the 
measurement
of parameters related to the strong phase of \dkpp amplitude performed 
by the CLEO collaboration. The analysis uses full data set of 
$772\times 10^6$ $B\overline{B}$ pairs collected by the Belle 
experiment at $\Upsilon(4S)$ resonance. 
We obtain $\phi_3 = (77.3^{+15.1}_{-14.9} \pm 4.2 \pm 4.3)^{\circ}$
and the suppressed amplitude ratio 
$r_B = 0.145\pm 0.030 \pm 0.011\pm 0.011$. Here the first error 
is statistical, the second is experimental systematic uncertainty, 
and the third is the error due to precision of strong phase 
parameters obtained by CLEO. This result is preliminary. 
\end{abstract}
\maketitle

\tighten

\section{Introduction}

The angle $\phi_3$ (also denoted as $\gamma$) is one of the least 
well-constrained parameters of the Unitarity Triangle. 
The theoretical uncertainties in $\phi_3$ determination are expected 
to be negligible, and the main difficulty in its measurement 
is a very low probability of the decays involved. 

The measurement that currently dominates the $\phi_3$ sensitivity uses 
$\bdk$ decay with the neutral $D$ meson decaying to 3-body final 
state such as $\kspp$~\cite{giri, binp_dalitz}. 
The weak phase $\phi_3$ appears in the 
interference between $b\to c\bar{u}s$ and $b\to u\bar{c}s$ transitions
and can be measured in the Dalitz plot analysis of the $D$ decay. 
This method requires the knowledge of the amplitude of \dnkpp decay
including its complex phase. The amplitude is obtained from the 
model that involves isobar and K-matrix descriptions of the decay 
dynamics, and thus results in the model uncertainty of the $\phi_3$
measurement. In the latest model-dependent Dalitz plot analyses 
performed by BaBar and Belle, this uncertainty ranges from 3$^{\circ}$ 
to 9$^{\circ}$~\cite{babar_gamma_1, babar_gamma_2, babar_gamma_3, belle_phi3_1, belle_phi3_2, belle_phi3_3}. 

The method to eliminate the model uncertainty using the 
binned Dalitz plot analysis has been proposed
by Giri {\it et al.}~\cite{giri}. The information about the strong 
phase in \dnkpp decay is obtained from the decays of quantum-correlated
$D^0$ pairs produced in $\psi(3770)\to D^0\overline{D}{}^0$ process. 
As a result, the model uncertainty is substituted by the statistical 
error related to the precision of strong phase parameters. 
The method has been further developed in~\cite{modind2006, modind2008}
where the experimental feasibility of the method has been shown 
and the analysis procedure has been proposed to optimally 
use the available $B$ decays and correlated $D^0$ pairs. 
Recently, the measurement of strong phase in \dnkpp and $D^0\to K^0_SK^+K^-$ 
decays has been performed by CLEO collaboration~\cite{cleo_1, cleo_2}. 
In this paper, we report the first measurement of $\phi_3$ 
with the model-independent Dalitz plot analysis of \dkpp 
decay from the mode \bdk, based on a 711 fb$^{-1}$ data sample
(corresponding to $772\times 10^6$ $B\overline{B}$ pairs)
collected by the Belle detector at the KEKB asymmetric 
$e^+e^-$ factory using the CLEO measurement. These results are preliminary. 

\section{The model-independent Dalitz plot analysis technique}

The amplitude of the \bdkp, \dkpp decay is given by the interference 
of $B^+\to \overline{D}{}^0K^+$ and $B^+\to D^0K^+$ amplitudes: 
\begin{equation}
  \ab=\adbar+ r_Be^{i(\delta_B+\phi_3)}\ad\,, 
\end{equation}
where $\adbar=\adbar(m^2_{K_S\pi^+}, m^2_{K_S\pi^-})\equiv\adbar\dvar$ 
is the amplitude of the \dbkpp decay,  $\ad=\ad\dvar$ 
is the amplitude of the \dnkpp decay ($\ad\dvar = \adbar(m^2_{-},m^2_{+})$ 
in the case of $CP$ conservation in $D$ decay), 
$r_B$ is the ratio of the absolute values of the two interfering 
amplitudes, and $\delta_B$ is the strong phase difference between 
them. The Dalitz plot density of the $D$ decay from \bdkp\ 
is given by 
\begin{equation}
 \begin{split}
  P_{B}=|\ab|^2 = & |\adbar+ r_Be^{i(\delta_B+\phi_3)}\ad|^2 = \\
  & \pdbar +r_B^2\pd + 2\sqrt{\pd\pdbar}(x_+ C+y_+ S)\,, 
 \label{p_b}
 \end{split}
\end{equation}
where
\begin{equation}
  x_+ = r_B\cos(\delta_B+\phi_3)\,; \;\;\;
  y_+ = r_B\sin(\delta_B+\phi_3)\,.
\end{equation}
The functions $C=C\dvar$ and $S=S\dvar$ are the cosine and sine of the strong 
phase difference $\delta_D=\arg\adbar-\arg\ad$ between the \dbkpp and \dnkpp 
amplitudes\footnote{
  This paper follows the convention for strong phases in $D$ decay
  amplitudes introduced in Ref.~\cite{modind2008}. 
}: 
\begin{equation}
  C=\cos\delta_D(m^2_+,m^2_-)\,; \;\;\;
  S=\sin\delta_D(m^2_+,m^2_-)\,. 
\end{equation}
The equations for the charge-conjugate mode \bdkm are obtained 
with the substitution $\phi_3 \rightarrow -\phi_3$ and 
$A\leftrightarrow \overline{A}$; the corresponding parameters 
of the admixture of the suppressed amplitude are 
\begin{equation}
  x_- = r_B\cos(\delta_B-\phi_3)\,; \;\;\;
  y_- = r_B\sin(\delta_B-\phi_3)\,.
\end{equation}
Using both $B$ charges, one can obtain $\phi_3$ and $\delta_B$ separately. 

Up to this point, the description of the model-dependent and 
model-independent techniques is the same. The model-dependent analysis deals
directly with the Dalitz plot density, and the functions $C$ and $S$ 
are obtained from model assumptions in the fit of the \dnkpp amplitude. 
In the model-independent approach, the Dalitz plot is divided into 
$2\mathcal{N}$ bins symmetric under the exchange $m^2_-\leftrightarrow m^2_+$. 
The bin index ``$i$'' ranges from $-\mathcal{N}$ to $\mathcal{N}$ (excluding 0); 
the exchange $m^2_+ \leftrightarrow m^2_-$ corresponds to the exchange 
$i\leftrightarrow -i$. The expected number of events in the bin ``$i$'' 
of the Dalitz plot of $D$ from \bdkp is
\begin{equation}
  N^{+}_i = 
  h_{B} \left[
    K_i + r_B^2K_{-i} + 2\sqrt{K_iK_{-i}}(x_+ c_i + y_+ s_i)
  \right] \,, 
  \label{n_b}
\end{equation}
where $h_{B}$ is a normalization constant and $K_i$ is the number of events 
in the $i$-th bin of the Dalitz plot of the $D$ meson in a flavor eigenstate 
(obtained using $D^{*\pm}\to D\pi^\pm$ sample). 
The terms $c_i$ and $s_i$ include information about 
the cosine and sine of the phase difference averaged over the bin region:
\begin{equation}
  c_i=\frac{\int\limits_{\mathcal{D}_i}
            \aad\aadbar
            \cos\delta_D\,d\mathcal{D}
            }{\sqrt{
            \int\limits_{\mathcal{D}_i}\aad^2 d\mathcal{D}
            \int\limits_{\mathcal{D}_i}\aadbar^2 d\mathcal{D}
            }}\,. 
  \label{cs}
\end{equation}
Here $\mathcal{D}$ represents the Dalitz plot phase space and 
$\mathcal{D}_i$ is the bin region over which the integration is performed. 
The terms $s_i$ are defined similarly with cosine substituted by sine. 

The absence of $CP$ violation in $D$ decay requires $c_i = c_{-i}$ and
$s_i = - s_{-i}$. The values of $c_i$ and $s_i$ terms can be measured 
in the quantum correlations of $D$ pairs by charm-factory experiments 
operated at the threshold of $D\overline{D}$ pair 
production~\cite{cleo_1, cleo_2}. 
The wave function of the two mesons is antisymmetric, thus 
the four-dimensional density of two correlated \dkpp Dalitz plots is
\begin{equation}
\begin{split}
  |A_{\rm corr}&(m_+^2,m_-^2,m'^2_+,m'^2_-)|^2=|\ad_1\adbar_2-\adbar_1\ad_2|^2=\\
     &\pd_1\pdbar_2 + \pdbar_1 \pd_2 -
     2\sqrt{\pd_1\pdbar_2\pdbar_1\pd_2}(C_1 C_2+S_1 S_2)\,, 
  \label{p_corr}
\end{split}
\end{equation}
where the indices ``1'' and ``2'' correspond to the two decaying $D$ mesons. 
In the case of a binned analysis, the number of events in the region of the
$(K^0_S\pi^+\pi^-)^2$ phase space described by the indices ``$i$'' and ``$j$'' is 
\begin{equation}
\begin{split}
  M_{ij} = &K_i K_{-j} + K_{-i} K_j  \\
    -&2\sqrt{K_iK_{-i}K_jK_{-j}}(c_i c_j + s_i s_j). 
\end{split}
\end{equation}
Another possibility is to use decays of $D$ in a $CP$-eigenstate to 
$K^0_S\pi^+\pi^-$ tagged by the other $D$ decaying to a $CP$-odd or 
$CP$-even final state. This allows to obtain constraints on the value 
of $c_i$. 

Once the values of the terms $c_i$ and $s_i$ are known, 
the system of equations (\ref{n_b})
contains only three free parameters ($x$, $y$, and $h_B$)
for each $B$ charge, and can be solved using a maximum likelihood 
method to extract the value of $\phi_3$. 

We have neglected charm mixing effects in decays of $D$ from 
both the \bdk process and in quantum-correlated $D\overline{D}$ production. 
It has been shown~\cite{modind_mixing} that although charm 
mixing correction is of the first order in the mixing parameters $x, y$, 
it is numerically small (of the order $0.2^{\circ}$ for $x,y\sim 0.01$) 
and can be neglected at the current level of precision. 
The future precision measurements of $\phi_3$
can account for both charm mixing and $CP$ violation (both in mixing 
and decay) once the corresponding parameters are measured.

Note that technically the system (\ref{n_b}) can be solved without 
external constraints on $c_i$ and $s_i$ for $\mathcal{N} \ge 2$.
However, due to the small value of $r_B$, there is very little
sensitivity to the $c_i$ and $s_i$ parameters in \bdk decays, which 
results in a reduction in the precision on $\phi_3$ that can be 
obtained~\cite{modind2006}.

\section{CLEO input}

The procedure of binned Dalitz plot analysis should give the correct 
results for any binning. However the statistical accuracy depends strongly 
on the amplitude behavior across the bins. Significant variations of 
the amplitude within a bin results in loss of coherence in the interference term. 
This effect becomes especially significant with limited 
statistics when a small number of bins has to be used. Better statistical 
precision is reached for the binning with the phase difference 
between the \dn and \dnbar amplitudes varying 
as little as possible~\cite{modind2008}. For the optimal precision, 
one also has to take the variations of the absolute value of the amplitude 
into account and the contribution of background events. 
The procedure to optimize the binning from the point of view of $\phi_3$ 
statistical precision has been proposed in~\cite{modind2008} and 
generalized to the case with the background in~\cite{cleo_2}. 
It has been shown that as little as 16 bins are enough 
to reach the statistical precision just 10--20\% smaller than in the 
unbinned case. 

\begin{figure}
  \includegraphics[width=0.47\textwidth]{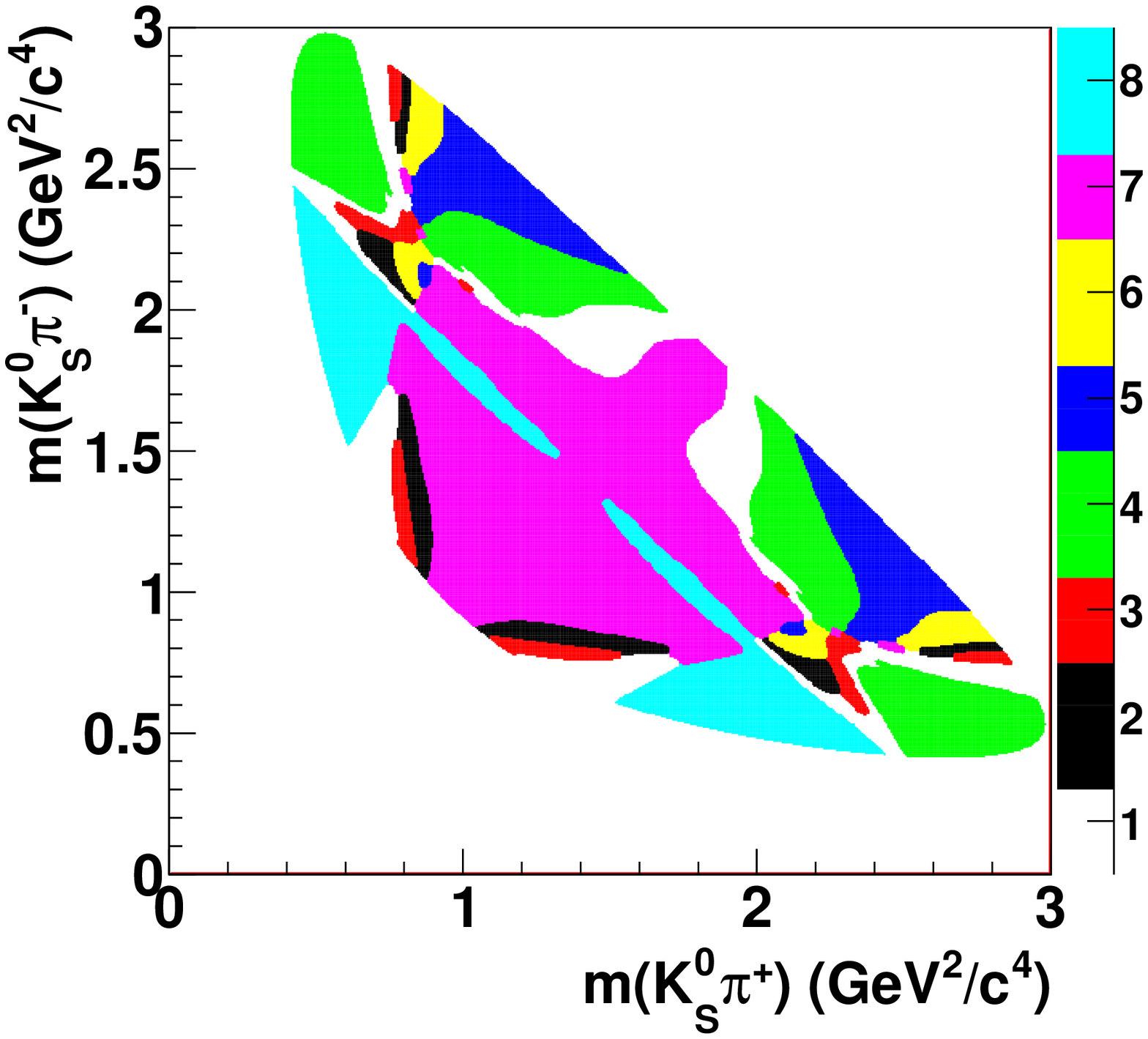}
  \caption{Optimal binning of \dkpp Dalitz plot. }
  \label{fig:babar_opt_bins}
\end{figure}

The optimization of binning sensitivity uses the amplitude of \dkpp 
decay. It should be noted, however, that although the choice of 
binning is model-dependent, a poor choice of model results only in the loss 
of precision but not in the bias of measured parameters. 
CLEO measured $c_i,s_i$ terms for four different binnings with 16 bins 
(bin index $i\in (-8,{\dots} -1, 1,{\dots} 8)$): 
\begin{enumerate}
  \item The binning optimized for statistical precision according 
        to the procedure from~\cite{modind2008} 
        (see Fig.~\ref{fig:babar_opt_bins}). The effect of the 
        background is not taken into account. 
        The amplitude is taken from the BaBar measurement~\cite{babar_gamma_2}. 
  \item Same as above, but optimized for the analysis with high 
        background in $B$ data ({\it e.~g.} at LHCb)
  \item The binning with bins equally distributed in the phase 
        difference $\Delta\delta_D$ between the \dn and \dnbar 
        decay amplitudes, with the amplitude from the BaBar 
        measurement~\cite{babar_gamma_2}. 
  \item Same as above, but with the amplitude 
        from the Belle analysis~\cite{belle_phi3_3}. 
\end{enumerate}

Our analysis uses the optimal binning shown in Fig.~\ref{fig:babar_opt_bins}
(option 1) as the baseline since it offers better statistical accuracy. 
In addition, we use the equal phase difference binning (\ddd-binning, 
option 3) as a cross-check. 

\begin{figure}
  \begin{center}
  \includegraphics[width=0.37\textwidth]{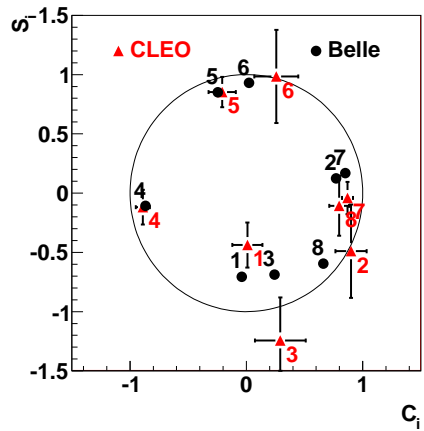}
  \end{center}
  \caption{Comparison of phase terms $c_i, s_i$ for the optimal binning
           measured by CLEO, and calculated from the Belle \dkpp amplitude model. }
  \label{fig:cs_cleo_opt}
\end{figure}

The results of the CLEO measurement of $c_i,s_i$ terms for the optimal 
binning are presented in Table~\ref{tab:cs_cleo_opt}. The same results 
in graphical form are shown in Fig.~\ref{fig:cs_cleo_opt}. 
The comparison with the $c_i,s_i$ calculated from the Belle 
model~\cite{belle_phi3_3} is presented, and shows a reasonable agreement 
between the Belle model and measurement. 

\begin{table}
  \caption{Values of the phase terms $c_i, s_i$ for the optimal 
           binning measured by CLEO~\cite{cleo_2}, 
           and calculated from Belle \dkpp amplitude model. 
           The $\chi^2/ndf$ of the agreement between the measured and 
           predicted $c_i,s_i$ is 18.6/16. }
  \label{tab:cs_cleo_opt}
  \begin{tabular}{|l|c|c|}
    \hline
                 &           CLEO measurement & Belle model    \\
    \hline
    $c_1$        & $-0.009\pm 0.088\pm 0.094$ & $-0.039$         \\
    $c_2$        & $+0.900\pm 0.106\pm 0.082$ & $+0.771$         \\
    $c_3$        & $+0.292\pm 0.168\pm 0.139$ & $+0.242$         \\
    $c_4$        & $-0.890\pm 0.041\pm 0.044$ & $-0.867$         \\
    $c_5$        & $-0.208\pm 0.085\pm 0.080$ & $-0.246$         \\
    $c_6$        & $+0.258\pm 0.155\pm 0.108$ & $+0.023$         \\
    $c_7$        & $+0.869\pm 0.034\pm 0.033$ & $+0.851$         \\
    $c_8$        & $+0.798\pm 0.070\pm 0.047$ & $+0.662$         \\
    \hline
    $s_1$        & $-0.438\pm 0.184\pm 0.045$ & $-0.706$         \\
    $s_2$        & $-0.490\pm 0.295\pm 0.261$ & $+0.124$         \\
    $s_3$        & $-1.243\pm 0.341\pm 0.123$ & $-0.687$         \\
    $s_4$        & $-0.119\pm 0.141\pm 0.038$ & $-0.108$         \\
    $s_5$        & $+0.853\pm 0.123\pm 0.035$ & $+0.851$         \\
    $s_6$        & $+0.984\pm 0.357\pm 0.165$ & $+0.930$         \\
    $s_7$        & $-0.041\pm 0.132\pm 0.034$ & $+0.169$         \\
    $s_8$        & $-0.107\pm 0.240\pm 0.080$ & $-0.596$         \\
    \hline
  \end{tabular}
\end{table}

\section{Analysis procedure}

The key equation of the analysis (\ref{n_b}) holds in the ideal situation with 
no background, constant efficiency (independent of Dalitz plot variables), 
and no cross-feed between the bins due to momentum resolution and radiative 
corrections. In this section, we outline the procedure of the analysis 
taking the above mentioned effects into account. 

\subsection{Efficiency profile}

\label{sec:eff}

The effect of non-uniform efficiency over the Dalitz plot is canceled out 
by using the flavor-tagged $D$ sample with the same kinematic properties 
as the sample for the signal $B$ decay. This approach allows of removal of the 
systematic error associated with the possible inaccuracy in the description 
of the detector acceptance in the Monte-Carlo (MC) simulation. 

We note that the Equations (\ref{p_b}) and (\ref{p_corr}) do not change 
after the transformation $p\to \epsilon p$, if the efficiency profile 
$\epsilon(m^2_+, m^2_-)$ is symmetric: 
$\epsilon(m^2_+, m^2_-)=\epsilon(m^2_-, m^2_+)$. It means that if the 
efficiency profile is the same in all three measurements involved 
(flavor $D$, correlated $\psi(3770)\to D\overline{D}$, and $D$ from $B\to DK$), 
the resulting measurement will be unbiased even if no efficiency correction 
is applied. 

We match the Dalitz plot efficiency profiles for the flavor $D$ to the 
one from $B\to DK$ by taking the flavor-tagged $D$ mesons with the same 
average momentum as the $D$ mesons from $B\to DK$. The center-of-mass (CM) 
$D$ momentum 
distribution for $B\to DK$ decays is practically uniform in the narrow 
range $2.1\mbox{ GeV}/c < p_D < 2.45\mbox{ GeV}/c$. We assume that 
the efficiency profile depends mostly on the $D$ momentum and take the 
flavor-tagged sample with the average momentum of $p_D=2.3$ GeV/$c$
(we use a wider range of $D$ momenta than in $B\to DK$ to increase the 
statistics). The assumption that the efficiency profile depends only 
on the $D$ momentum is tested using MC simulation, and the remaining 
difference is treated as the systematic uncertainty. 

While calculating $c_i, s_i$, CLEO applies an efficiency correction, 
therefore the values reported in their analysis correspond to flat 
efficiency profile. 
To use the $c_i, s_i$ values in the $\phi_3$ analysis, they have to be 
corrected for the Belle efficiency profile. This correction cannot be 
performed in a completely model-independent way, since the correction terms 
include the phase variation inside the bin. Fortunately, the calculations
using the Belle \dkpp model show that this correction is negligible even 
for very large non-uniformity of the efficiency profile. The difference 
between the uncorrected $c_i,s_i$ terms and those corrected for the 
efficiency, calculated using the efficiency profile parametrization 
used in the 605 fb$^{-1}$ analysis~\cite{belle_phi3_3}, 
does not exceed 0.01, {\it i.~e.} it is negligible compared to the 
statistical error.  

\subsection{Momentum resolution}

Finite momentum resolution leads to migration of events between the bins. 
In the binned approach, this effect can be corrected for in a 
non-parametric way. The migration can be described by the linear transformation 
of the number of events in bins:
\begin{equation}
N'_i = \sum \alpha_{ik} N_k, 
\end{equation}
where $N_k$ is the number of events the bin $k$ would contain without the 
cross-feed, and $N'_i$ is the reconstructed number of events in the bin $i$. 
The cross-feed matrix $\alpha_{ik}$ is nearly unit: $\alpha_{ik} \ll 1$ for 
$i\neq k$. It is obtained from the signal MC simulation with the amplitude 
measured in Belle's 605 fb$^{-1}$ analysis~\cite{belle_phi3_3}. In the case 
of \dkpp decay from $B$, the cross-feed depends on the parameters $x,y$. 
We assume that this effect is small and neglect it. 

Migration of events between the bins occurs also due to final state 
radiation (FSR). The $c_i, s_i$ terms in the CLEO measurement are not 
corrected to FSR; we therefore do not simulate FSR to obtain the 
cross-feed matrix to minimize the bias due to this effect. Comparison of 
the cross-feed with and without FSR shows that this effect is negligible. 

\subsection{Fit procedure}

\label{sec:fit_procedure}

The background contribution has to be accounted for in the calculation 
of the values $N_i$ and $K_i$. Statistically the most effective way of 
calculating the number of signal events (especially in the case of $N_i$, 
where the statistics is a limiting factor) is to perform the unbinned fit 
in the event selection variables for the events in each bin ``$i$'' of the 
Dalitz plot. 

Two different approaches are used in this analysis. In the first one, 
we fit the data distribution in each bin separately, with the 
number of events for signal and backgrounds as free parameters. 
Once the numbers of events in bins $N_i$ are found, we 
use them in Eq.~\ref{n_b} to obtain the parameters $x_{\pm}, y_{\pm}$.
Technically it is done by minimizing the 
negative logarithmic likelihood of the form
\begin{equation}
          -2\log\mathcal{L}(x,y)=
             -2\sum_i\log p(\langle N_i\rangle(x,y),N_i,\sigma_{N_i}),
\end{equation}
where $\langle N_i\rangle(x,y)$ is the expected number of events in 
the bin $i$ obtained from Eq.~\ref{n_b}, $N_i$ and $\sigma_{N_i}$
are the observed number of events and its error obtained from 
the data fit. If the probability density function (PDF) $p$ is Gaussian, 
this procedure translates to the $\chi^2$ fit. 

The procedure described above does not make any assumptions on the 
Dalitz distribution of the background events, since the fits in each bin are
independent. Thus there is no associated systematic uncertainty. However, 
in the case of low number of events and many background components this 
can be a limiting factor. Another solution is to use the combined 
fit with a common likelihood for all bins. Relative numbers of background
events in bins in such a fit can be constrained externally from {\it e.~g.}
MC sample. In addition, in the case of the combined fit, the 
two-step procedure of first extracting the numbers of signal events, and 
then using them to obtain $(x,y)$ is not needed --- the expected numbers 
of events $\langle N_i\rangle$ as functions of $(x,y)$ can be plugged 
directly into the likelihood. Thus the variables $(x,y)$ become free 
parameters of the combined likelihood fit, and the assumption of the 
Gaussian distribution of the number of signal events is not needed. 

Both approaches (separate fits in bins, and the combined fit) are tested with  
the control sample and the MC simulation. We choose the combined fit 
approach as the baseline, but the approach with separate fits in bins is 
also used: it allows to clearly demonstrate the $CP$ asymmetry of the 
number of events in bins. 

\section{Event selection}

We use a data sample of $772\times 10^6$ $B\overline{B}$ pairs 
collected by the Belle detector. The decay chains \bdkp\ and \bdpi\ 
are selected for the analysis. The neutral $D$ meson is reconstructed 
in the $K^0_S\pi^+\pi^-$ final state in all cases. We also select decays 
of \dsdpi\ produced via the $e^+e^-\to c\bar{c}$ continuum process as 
a high-statistics sample to determine the terms related to flavor-tagged 
\dnkpp decay. 

The Belle detector is described in detail elsewhere \cite{belle,svd2}. 
It is a large-solid-angle magnetic spectrometer consisting of a
silicon vertex detector (SVD), a 50-layer central drift chamber (CDC) for
charged particle tracking and specific ionization measurement ($dE/dx$), 
an array of aerogel threshold Cherenkov counters (ACC), time-of-flight
scintillation counters (TOF), and an array of CsI(Tl) crystals for 
electromagnetic calorimetry (ECL) located inside a superconducting solenoid coil
that provides a 1.5 T magnetic field. An iron flux return located outside 
the coil is instrumented to detect $K_L$ mesons and identify muons (KLM).

Charged tracks are required to satisfy criteria based on the quality of the 
track fit and the distance from the interaction point (IP). We require each track 
to have a transverse momentum greater than 100 MeV/$c$, and the impact 
parameter relative to the IP of the beams less than 2 mm in 
the transverse and less than 10 mm in longitudinal projections. 
Separation of kaons and pions is accomplished by combining the responses of 
the ACC and the TOF with the $dE/dx$ measurement from the CDC. 
Neutral kaons are reconstructed from pairs of oppositely charged tracks
with an invariant mass $M_{\pi\pi}$ within $7$ MeV/$c^2$ of the nominal 
$K^0_S$ mass, flight distance from the IP in the plane transverse to 
the beam axis grater than 0.1 mm, 
and the cosine of the angle between the projections of flight direction 
and its momentum greater than 0.95. 

To determine the terms for flavor-tagged \dnkpp\ decay we use 
$D^{*\pm}$ mesons produced via the $e^+ e^-\to c\bar{c}$ continuum process. 
The flavor of the neutral $D$ meson is tagged by the charge of the slow 
pion in the decay \dsdpi. The slow pion track is required to originate 
from the $D^0$ decay vertex to improve the momentum and angular resolution. 
The selection of signal candidates is based on two variables, 
the invariant mass of the neutral $D$ candidates $M_D=M_{K^0_S\pi^+\pi^-}$ 
and the difference of the invariant masses of the $D^{*\pm}$ and the 
neutral $D$ candidates 
$\Delta M=M_{(K^0_S\pi^+\pi^-)_D\pi}-M_{K^0_S\pi^+\pi^-}$. 
We retain the events satisfying the following criteria: 
$1800\mbox{ MeV}/c^2<M_D<1920$ MeV/$c^2$ and $\Delta M<0.15$ MeV/$c^2$.
We also require the momentum of the $D^0$ candidate in the CM frame 
$p_D$ to be greater than 1.5~GeV/$c$. 
About 15\% of selected events contain more than one $D^{*\pm}$ candidate
that satisfies the requirements above; in that case we keep only one 
randomly selected candidate. 

Selection of \bdk and \bdpi samples is based on the CM energy difference
$\de = \sum E_i - E_{\rm beam}$ and the beam-constrained $B$ meson mass
$\mbc = \sqrt{E_{\rm beam}^2 - (\sum \vec{p}_i)^2}$, where $E_{\rm beam}$ 
is the CM beam energy, and $E_i$ and $\vec{p}_i$ are the CM energies and 
momenta of the $B$ candidate decay products. We select events with 
$\mbc>5.2$ GeV/$c^2$ and $|\de|<0.18$~GeV for further analysis. 
We also impose a requirement on the invariant mass of the neutral $D$ 
candidate $|M_{\kspp}-M_{D^0}|<11$~MeV/$c^2$. 

Further separation of the background from $e^+e^-\to q\bar{q}$ ($q=u, d, s, c$) 
continuum events is done by calculating two variables that characterize the 
event shape. One is the cosine of the thrust angle \thr, 
where $\theta_{\rm thr}$ is the angle between the thrust axis of 
the $B$ candidate daughters and that of the rest of the event, 
calculated in the CM frame. 
The other is a Fisher discriminant \fish composed of 11 parameters \cite{fisher}: 
the production angle of the $B$ candidate, the angle of the $B$ thrust 
axis relative to the beam axis, and nine parameters representing 
the momentum flow in the event relative to the $B$ thrust axis in the CM frame.

In both flavor $D^0$ and \bdk (\bdpi) samples the momenta of the tracks 
forming a $D^0$ candidate are fitted to the nominal $D^0$ mass in 
the calculation of the Dalitz plot variables. 

\section{Flavor-tagged sample \dsdpi, \dkpp}

\label{sec:flavor}

The numbers of events $K_i$ in bins of the flavor-tagged \dkpp decay are 
obtained from the two-dimensional unbinned fit of the distribution 
of variables $M_{D}$ and $\Delta M$. The fits in each Dalitz plot bin 
are done independently. We take the candidates in the CM $D$ momentum range
$1.8\mbox{ GeV}/c<p_D<2.8\mbox{ GeV}/c$. It provides the same average 
$p_D$ as in $B\to DK$ decays ($p_D=2.3$~GeV/$c$) to reduce the influence 
of the efficiency profile on $\phi_3$ measurement (see Section~\ref{sec:eff}). 

The fit uses the signal PDF and two background components: purely 
random combinatorial background
and the background with real $D^0$ and random slow pion track. 
The signal distribution is a product of the PDF's for $M_D$ (triple Gaussian) and 
$\Delta M$ (sum of bifurcated Student distribution and bifurcated 
Gaussian distribution). The combinatorial background is parametrized 
by the linear function in $M_D$ and by the function 
with a kinematic threshold at the $\pi^+$ mass in $\Delta M$. 
The random slow pion background is parametrized as a 
product of signal $M_D$ distribution and the combinatorial $\Delta M$ 
background shape. 

The parameters of the signal and background distributions are obtained 
from the data fit. The parameters of the signal PDF are constrained 
to be the same in all bins. The free parameters in each bin are 
the number of signal events $K_i$, the parameters of the 
background distribution and fractions of the background components. 

\begin{figure}
  \includegraphics[width=0.5\textwidth]{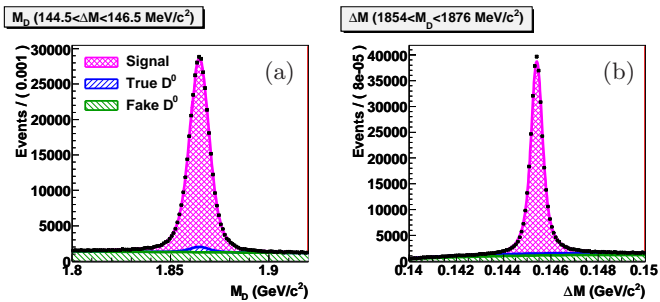}
  \put(-153,90){(a)}
  \put(-25,90){(b)}
  \caption{Projections of the flavor-tagged \dsdpi, \dkpp data with 
           $1.8\mbox{ GeV}/c<p_D<2.8\mbox{ GeV}/c$ onto 
           (a) $M_D$ and (b) $\Delta M$ variables. 
           Histograms show fitted signal and background 
           contributions, points with the error bars are the data.
           Full \dkpp Dalitz plot is used. }
  \label{fig:dsdpi_dmmd}
\end{figure}

The fit results of the flavor sample for the whole Dalitz plot are 
shown in Fig.~\ref{fig:dsdpi_dmmd}. The number of signal events 
calculated from the integral of the signal distribution is 
$426938\pm 825$, the background fraction in the signal box 
$|M_D-m_{D^0}|<11$ MeV/$c^2$, $144.5\mbox{ MeV}/c^2<\Delta M<146.5$ MeV/$c^2$
is $10.1\pm 0.1$\%. The numbers of events in bins are shown in 
Table~\ref{tab:flavor_bins}. 

\begin{table}
  \caption{Numbers of events in Dalitz plot bins for the flavor-tagged 
  \dsdpi, \dkpp sample with $1.8\mbox{ GeV}/c<p_D<2.8\mbox{ GeV}/c$. 
  Results of the 2D $(\Delta M, M_D)$ fit to data. }
  \label{tab:flavor_bins}
  \begin{tabular}{|r|c|}
   \hline
    Bin $i$  & $K_i(p_D\in(1.8, 2.8))$ \\
   \hline
   -8 & $26450\pm 181$             \\
   -7 & $22476\pm 196$             \\
   -6 & $1765\pm 68$               \\
   -5 & $13146\pm 143$             \\
   -4 & $26482\pm 202$             \\
   -3 & $1601\pm 58$               \\
   -2 & $1827\pm 63$               \\
   -1 & \phantom{$0$}$8770\pm 124$ \\
    1 & $43261\pm 255$             \\
    2 & $58005\pm 268$             \\
    3 & $62808\pm 274$             \\
    4 & $44513\pm 253$             \\
    5 & $21886\pm 177$             \\
    6 & $28876\pm 197$             \\
    7 & $48001\pm 265$             \\
    8 & \phantom{$0$}$9279\pm 125$ \\
   \hline
    Total & $426938\pm 825$\phantom{$0$} \\
   \hline
  \end{tabular}
\end{table}

\section{Selection of \bdpi and \bdk samples}

The decays \bdk and \bdpi have similar topology and background 
sources and their selection is performed in a similar way. 
The mode \bdpi has an order of magnitude larger branching 
ratio and small amplitude ratio $r_B\sim 0.01$ due to small ratio 
of the weak coefficients 
$|V_{ub}^{\vphantom{*}} V^*_{cd}|/
 |V_{cb}^{\vphantom{*}} V^*_{ud}|\sim 0.02$ and additional color 
suppression factor as in the case of \bdk. This mode is used as 
a control sample to test the procedures of the background extraction
and Dalitz plot fit. Also, the resolution scale factors and 
Dalitz plot structure of some background components are constrained 
from the control sample and used in the signal fit. 

Extraction of the number of signal events is performed by fitting 
the 4D distribution of variables \mbc, \de, \thr and \fish. 
The fit to \bdpi sample uses three background components in addition to the 
signal PDF. These are: 
\begin{itemize}
  \item Combinatorial background from $e^+e^-\to q\overline{q}$ process, 
        where $q=(u,d,s,c)$. 
  \item Random $B\overline{B}$ background, where the tracks forming the 
        \bdpi candidate come from decays of both $B$ mesons in the event. 
        The number of possible $B$ decay combinations that contribute 
        to this background is large, therefore both the Dalitz 
        distribution and $(\mbc, \de)$ distribution are quite smooth. 
  \item Peaking $B\overline{B}$ background, where all tracks forming 
        the \bdpi candidate come from the same $B$ meson.  This kind of 
        background is dominated by the $B\to D^*\pi$ decays with 
        lost $\pi$ or $\gamma$ from $D^{*}$ decay. 
\end{itemize}
The \bdk fit includes in addition the background from \bdpi decays with 
pion misidentified as kaon. 

The PDF for the signal parametrization (as well as for each of the 
background components) is a product of $(\mbc,\de)$ and $(\thr,\fish)$
PDFs. The $(\mbc,\de)$ PDF is a sum of two 2D Gaussian distributions
(core and wide) with the correlation between $\mbc$ and $\de$. 
We use common parametrization for the $\thr,\fish$ distribution 
for signal and all background components. The distribution is 
parametrized by the sum of two functions (with different coefficients) 
of the form
\begin{equation}
  \begin{split}
  p(x,\fish) = & \exp(C_1x+C_2x^2+C_3x^3)\times \\
               & G(\fish, F_0(x), \sigma_{FL}(x), \sigma_{FR}(x)), 
  \end{split}
  \label{thrfish_param}
\end{equation}
where $x=\thr$, $G(x,F,\sigma_R,\sigma_R)$ is the bifurcated 
Gaussian distribution with the mean $F$ and the widths $\sigma_L$
and $\sigma_R$, and functions $F_0$, $\sigma_{FL}$ and $\sigma_{FR}$
are the polynomials which contain only even powers of $x$. 

Combinatorial background from continuum $e^+e^-\to q\overline{q}$
production is obtained from the experimental sample collected at the 
CM energy below $\Upsilon(4S)$ resonance (off-resonance data). 
The parametrization in variables $(\thr,\fish)$ is the same as described
for the signal PDF. 
The parametrization in $(\mbc,\de)$ is the product of exponential 
distribution in $\Delta E$ and the empirical shape proposed by the 
Argus collaboration~\cite{argus} in $\mbc$: 
\begin{equation}
  p_{\rm comb}(\mbc, \de)=\exp(-\alpha\de)\mbc\sqrt{y}\exp(-cy), 
\end{equation}
where $y=1-\mbc/E_{\rm beam}$, and $E_{\rm beam}$ is the CM beam energy. 

Random $B\overline{B}$ background is obtained from generic MC sample. 
Generator information is used to select only the events where the candidate 
is formed from tracks coming from both $B$ mesons. The $(\mbc,\de)$
distribution of this background is parametrized by the sum of three 
components:
\begin{itemize}
  \item product of exponential (in $\de$) and Argus (in $\mbc$) 
  functions, as for continuum background, 
  \item product of exponential in $\de$ and bifurcated Gaussian 
  distribution in $\mbc$, where the mean of the Gaussian distribution 
  is linear as a function of $\de$. 
  \item two-dimensional Gaussian distribution in $\de, \mbc$ with 
  correlation (asymmetric in $\mbc$). 
\end{itemize}

Peaking $B\overline{B}$ backgrounds are parametrized by the same 
function as the random $B\overline{B}$ background. The background 
coming from $B^+B^-$ and $B^0\overline{B}{}^0$ decays is treated 
separately in $(\mbc,\de)$ variables, while the common 
$(\thr,\fish)$ distribution is used. In the case of \bdk fit, 
we take the \bdpi events with pion misidentified as kaon as a
separate background category. The distributions of $\mbc, \de$
and $\thr, \fish$ variables are parametrized in the same way as for
the signal events and are obtained from MC simulation. 

The Dalitz plot distributions of the background components are 
discussed in the next section. Note that the Dalitz distribution is 
described by the relative number of events in bins. The numbers of 
events in bins can be taken as free parameters in the fit, thus 
there will be no uncertainty due to Dalitz plot description of the 
background in such an approach. This procedure is justified for the 
background that is well separated from the signal (such as peaking 
$B\overline{B}$ background in the case of \bdpi), or if the background 
is constrained by much larger number of events than the signal (such 
as the continuum background). 

\begin{figure}
  \includegraphics[width=0.5\textwidth]{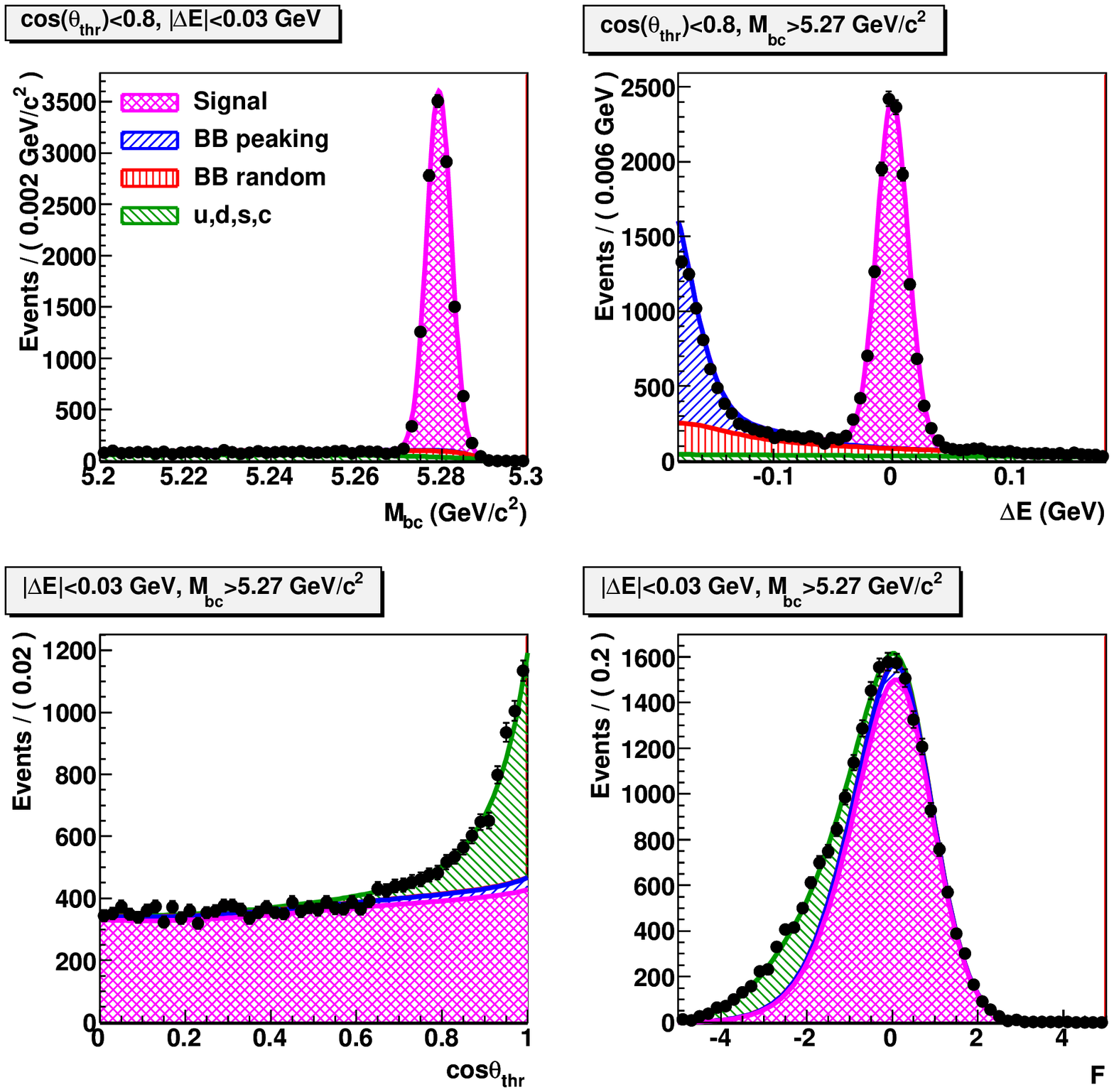}
  \put(-152,219){(a)}
  \put(-25,219){(b)}
  \put(-152,94){(c)}
  \put(-25,94){(d)}
  \caption{Projections of the \bdpi data onto 
           (a) \mbc, (b) \de, (c) \thr and (d) \fish variables. 
           Histograms show fitted signal and background 
           contributions, points with the error bars are the data. 
           Full \dkpp Dalitz plot is used. }
  \label{fig:bdpi_exp_all}
\end{figure}

\begin{figure}
  \includegraphics[width=0.5\textwidth]{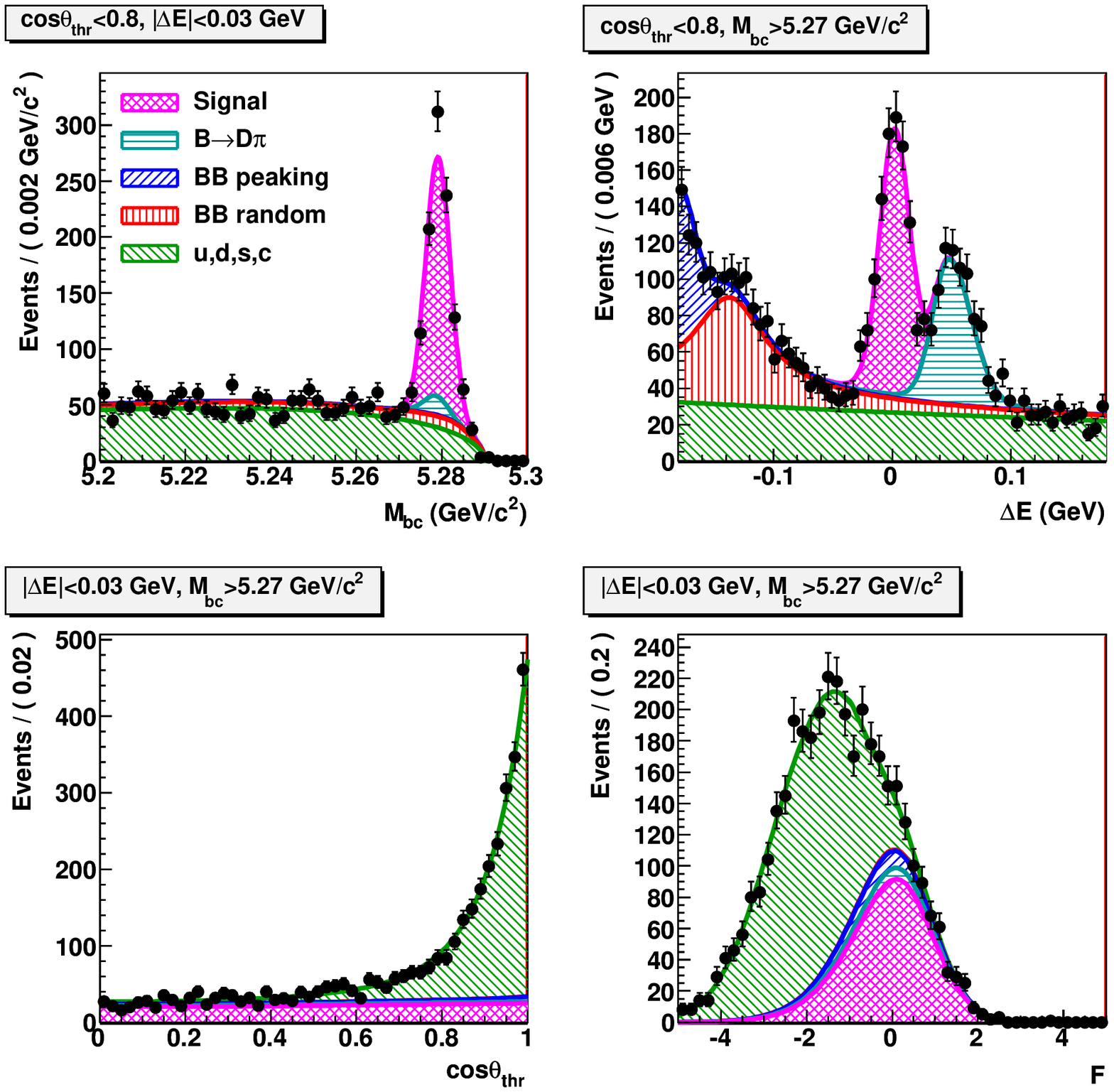}
  \put(-152,219){(a)}
  \put(-25,219){(b)}
  \put(-152,94){(c)}
  \put(-25,94){(d)}
  \caption{Projections of the \bdk data onto 
           (a) \mbc, (b) \de, (c) \thr and (d) \fish variables. 
           Histograms show fitted signal and background 
           contributions, points with the error bars are the data. 
           Full \dkpp Dalitz plot is used. }
  \label{fig:bdk_exp_all}
\end{figure}

The results of the fit to \bdpi and \bdk data with the full Dalitz plot 
taken are shown in Figs~\ref{fig:bdpi_exp_all} and \ref{fig:bdk_exp_all}, 
respectively. We obtain a total of $19106\pm 147$ signal \bdpi events and 
$1176\pm 43$ signal \bdk events --- 55\% more than in the 
605 fb$^{-1}$ model-dependent analysis~\cite{belle_phi3_3}. 
The improvement partially comes from larger integrated luminosity, and 
partially from the better selection efficiency due to improved tracking 
procedures. 

\section{Data fits in bins}

The data fits in bins for both \bdpi and \bdk are performed with two 
different procedures: separate fits for the number of events in bins, 
and the combined fit with the free parameters $(x,y)$, as discussed in 
Section~\ref{sec:fit_procedure}. The combined fit is used to obtain
the final values for $(x,y)$, while the separate fits provide the 
cross-check of the fit procedure and a way to visualize the 
$CP$-violating effect. The study with MC pseudo-experiments is performed 
to check that the observed difference in the fit results 
between the two approaches agrees with the expectation. 

In the case of separate fits in bins, we first perform the fit to all 
events in the Dalitz plot. The fit uses background shapes fixed from the 
generic MC simulation of continuum and $B\overline{B}$ decays. 
The signal shape is fixed from the signal MC sample except that we 
float the mean values of \de and \mbc and width scale factors. 
As a next step, we fit the 4D $(\mbc, \de, \thr, \fish)$ distributions
in each bin separately. The free parameters of each fit are the number of
signal events, and the number of events in each background 
category. 

\begin{figure}
  \includegraphics[width=0.5\textwidth]{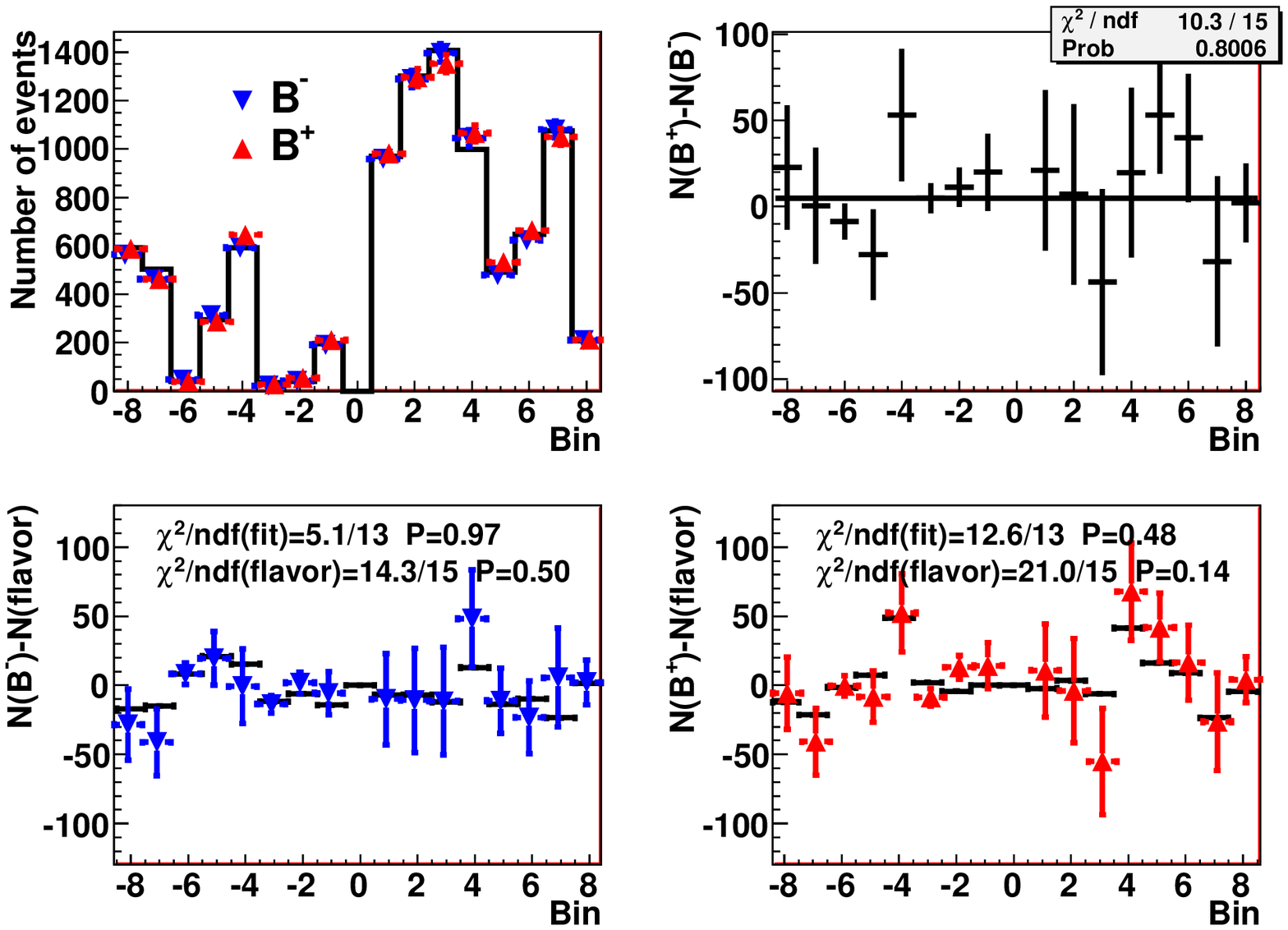}
  \put(-227,165){(a)}
  \put(-99,165){(b)}
  \put(-227,22){(c)}
  \put(-99,22){(d)}
  \caption{Results of the fit of \bdpi control sample. 
  (a) Numbers of events in bins of \dkpp Dalitz plot: from \bdpim (red), 
  \bdpip (blue) and flavor sample (histogram). 
  (b) Difference of the number of events from \bdpip and \bdpim decays. 
  (c) Difference of the number of events from \bdpim and flavor sample
  (normalized to the total number of \bdpim decays): data (points with 
  the error bars), and as a result of the $(x,y)$ fit (horizontal bars). 
  (d) Same for \bdpip data. }
  \label{fig:bdpi_bin_n}
\end{figure}

The numbers of signal events in bins extracted from the fits are given in 
Table~\ref{tab:bdpi_signal_num}. These numbers are used in the fit to extract 
$(x,y)$ using (\ref{n_b}) after the cross-feed and efficiency correction for 
both $N_i$ and $K_i$. Figure~\ref{fig:bdpi_bin_n} illustrates the results of
this fit. The numbers of signal events in bins separately for $B^+$ and $B^-$
are shown in Fig.~\ref{fig:bdpi_bin_n}(a) together with the numbers of events in the
flavor sample (appropriately scaled). The difference in the number of signal events 
shown in Fig.~\ref{fig:bdpi_bin_n}(b) does not reveal $CP$ violation. 
Figures~\ref{fig:bdpi_bin_n}(c) and (d) show the difference of the numbers of 
signal events for $B^+$ ($B^-$) data and scaled flavor sample, both for the 
data and after the $(x,y)$ fit. The $\chi^2/ndf$ is reasonable for both the 
$(x,y)$ fit and for the agreement with the purely flavor-specific amplitude. 

\begin{figure}
  \includegraphics[width=0.5\textwidth]{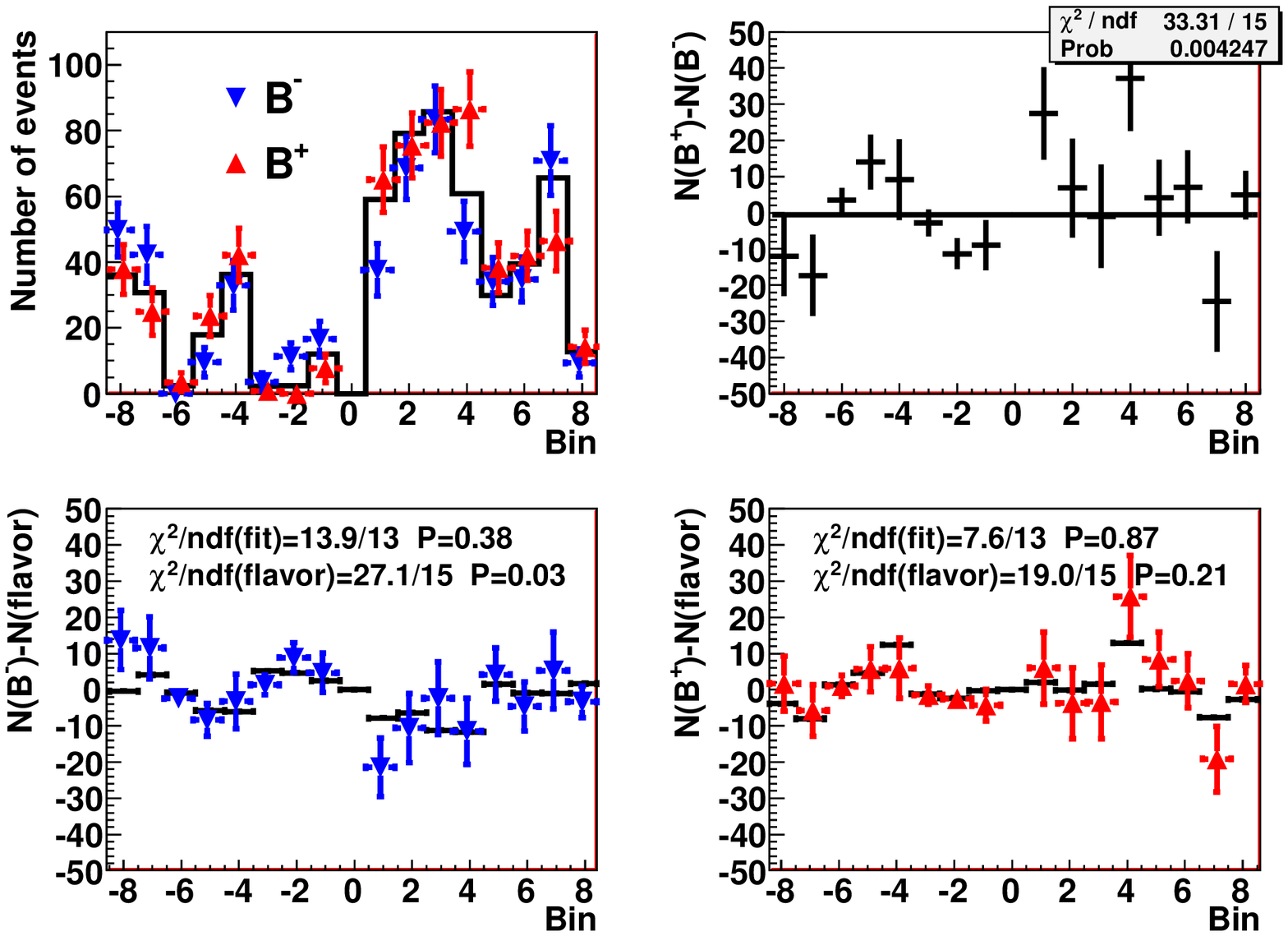}
  \put(-227,165){(a)}
  \put(-99,165){(b)}
  \put(-227,22){(c)}
  \put(-99,22){(d)}
  \caption{Results of the fit of \bdk control sample. 
  (a) Numbers of events in bins of \dkpp Dalitz plot: from \bdkm (red), 
  \bdkp (blue) and flavor sample (histogram). 
  (b) Difference of the number of events from \bdkp and \bdkm decays. 
  (c) Difference of the number of events from \bdkm and flavor sample
  (normalized to the total number of \bdkm decays): data (points with 
  the error bars), and as a result of the $(x,y)$ fit (horizontal bars). 
  (d) Same for \bdkp data. }
  \label{fig:bdk_bin_n}
\end{figure}

Unlike \bdpi, the \bdk sample shows significantly different 
numbers of events in bins of $B^+$ and $B^-$ data (see 
Fig.~\ref{fig:bdk_bin_n}(b) and Table~\ref{tab:bdk_signal_num}). 
The probability to obtain this difference 
as a result of statistical fluctuation is 0.42\%. This number can be 
taken as the model-independent measure of the $CP$ violation significance.
The significance of $\phi_3$ being nonzero is in general 
smaller since $\phi_3\neq 0$ results in a specific pattern of charge 
asymmetry. The fit of the numbers of events to the expected 
pattern described by the parameters $(x,y)$ shows a good 
quality~\ref{fig:bdk_bin_n}(c,d), {\it i. e.} is consistent 
with the hypothesis that the observed $CP$ violation is solely 
explained by the mechanism involving nonzero $\phi_3$. 

\begin{table}
  \caption{Numbers of events in Dalitz plot bins for the \bdpi, 
  \dkpp sample with the optimal binning. 
  Results of the independent 4D fits with variables $(\mbc, \de, \thr, \fish)$ fit to data. }
  \label{tab:bdpi_signal_num}
  \begin{tabular}{|r|c|c|}
   \hline
    Bin $i$  & $N^{-}_i$ & $N^{+}_i$ \\
   \hline
   -8     &  $564.2\pm 25.3$             &  $587.0\pm 25.7$             \\
   -7     &  $462.3\pm 23.8$             &  $462.8\pm 23.9$             \\
   -6     &   $47.9\pm 7.7$              &   $39.2\pm 7.2$              \\
   -5     &  $314.1\pm 19.0$             &  $286.2\pm 18.2$             \\
   -4     &  $592.6\pm 26.5$             &  $645.7\pm 27.8$             \\
   -3     &   $22.2\pm 6.2$              &   $27.2\pm 6.3$              \\
   -2     &   $42.7\pm 7.6$              &   $54.0\pm 8.7$              \\
   -1     &  $190.8\pm 15.4$             &  $210.8\pm 16.3$             \\
    1     &  $959.2\pm 32.6$             &  $980.2\pm 33.1$             \\
    2     & $1288.7\pm 37.0$\phantom{$0$}& $1295.9\pm 37.1$\phantom{$0$}\\
    3     & $1395.8\pm 38.4$\phantom{$0$}& $1352.2\pm 37.9$\phantom{$0$}\\
    4     & $1045.5\pm 34.7$\phantom{$0$}& $1065.1\pm 34.9$\phantom{$0$}\\
    5     &  $479.3\pm 23.3$             &  $532.2\pm 24.5$             \\
    6     &  $623.7\pm 26.0$             &  $663.5\pm 26.7$             \\
    7     & $1081.0\pm 35.3$\phantom{$0$}& $1049.2\pm 34.8$\phantom{$0$}\\
    8     &  $210.0\pm 16.1$             &  $212.1\pm 16.3$             \\
   \hline
    Total & $9467.1\pm 103.6$            & $9639.1\pm 104.7$            \\
   \hline
  \end{tabular}
\end{table}

\begin{table}
  \caption{Numbers of events in Dalitz plot bins for the \bdk, 
  \dkpp sample with the optimal binning. 
  Results of the independent 4D fits with variables $(\mbc, \de, \thr, \fish)$ fit to data. }
  \label{tab:bdk_signal_num}
  \begin{tabular}{|r|c|c|}
   \hline
    Bin $i$  & $N^{-}_i$ & $N^{+}_i$ \\
   \hline
   -8     &  $49.8\pm 8.2$              &  $37.8\pm 7.5$             \\
   -7     &  $42.2\pm 8.6$              &  $24.9\pm 7.2$             \\
   -6     &\phantom{$0$}$0.0\pm 1.9$    & \phantom{$0$}$3.4\pm 2.9$  \\
   -5     &\phantom{$0$}$9.6\pm 4.5$    &  $23.6\pm 6.2$             \\
   -4     &  $32.9\pm 7.5$              &  $42.1\pm 8.3$             \\
   -3     &\phantom{$0$}$3.5\pm 2.8$    & \phantom{$0$}$0.7\pm 2.5$  \\
   -2     &   $11.3\pm 4.1$             & \phantom{$0$}$0.0\pm 1.3$  \\
   -1     &  $16.6\pm 5.4$              & \phantom{$0$}$7.7\pm 4.4$  \\
    1     &  $37.6\pm 8.0$              &  $65.1\pm 9.9$             \\
    2     & $68.6\pm 9.6$               & $75.5\pm 9.8$              \\
    3     &\phantom{$0$}$83.4\pm 10.1$  & \phantom{$0$}$82.4\pm 10.2$ \\
    4     & $49.3\pm 9.1$               & \phantom{$0$}$86.5\pm 11.4$ \\
    5     &  $34.0\pm 7.3$              &  $38.3\pm 7.6$             \\
    6     &  $34.8\pm 6.8$              &  $41.9\pm 7.5$             \\
    7     &\phantom{$0$}$70.8\pm 10.6$  & $46.4\pm 9.0$              \\
    8     &\phantom{$0$}$9.4\pm 4.3$    &  $14.2\pm 5.1$             \\
   \hline
    Total & $574.9\pm 29.9$             & $601.6\pm 30.8$            \\
   \hline
  \end{tabular}
\end{table}

\begin{figure}
  \begin{center}
  \includegraphics[width=0.37\textwidth]{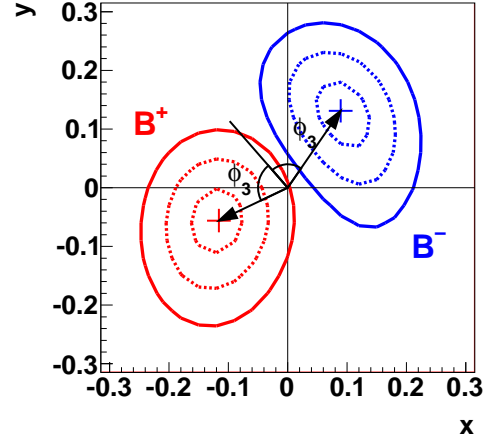}
  \end{center}
  \caption{One-, two-, and three standard deviations levels for $x,y$ fit
           of \bdk mode. }
  \label{fig:bdk_xy}
\end{figure}

The default combined fit uses the constraint of the random $B\overline{B}$
background in bins from the generic MC, and $x_{\pm},y_{\pm}$ variables as free 
parameters. Fits to $B^+$ and $B^-$ data are performed separately. 
Additional free parameters are the numbers of continuum and peaking 
$B\overline{B}$ backgrounds in each bin, fraction of the random $B\overline{B}$
background, and means and scale factors of the signal \mbc and \de distributions. 
The values of $x,y$ are then corrected for the fit bias obtained from 
MC pseudo-experiments. The value of the bias depends on the initial 
$x$ and $y$ values and is of the order $5\times 10^{-3}$ for \bdk sample 
and less than $10^{-3}$ for \bdpi sample. 

The values of $x,y$ parameters and their statistical correlations 
obtained from the combined fit are as follows:
\begin{equation}
\begin{split}
x_- &= -0.0045\pm 0.0087\pm 0.0050\pm 0.0026, \\
y_- &= -0.0231\pm 0.0107\pm 0.0050\pm 0.0065, \\
\rm{corr}(x_-, y_-) & = -0.189,\\
x_+ &= -0.0172\pm 0.0089\pm 0.0060\pm 0.0026, \\
y_+ &= +0.0129\pm 0.0103\pm 0.0060\pm 0.0065, \\
\rm{corr}(x_+, y_+) & = -0.205
\end{split}
\end{equation}
for \bdpi control sample and 
\begin{equation}
\begin{split}
x_- &= +0.095\pm 0.045\pm 0.014\pm 0.017, \\
y_- &= +0.137^{+0.053}_{-0.057}\pm 0.019\pm 0.029, \\
\rm{corr}(x_-, y_-) &= -0.315,\\
x_+ &= -0.110\pm 0.043\pm 0.014\pm 0.016,\\
y_+ &= -0.050^{+0.052}_{-0.055}\pm 0.011\pm 0.021,\\
\rm{corr}(x_+, y_+) & = +0.059
\end{split}
\end{equation}
for \bdk sample. Here the first error is statistical, 
the second error is the systematic uncertainty, and the 
third error is the uncertainty due to the errors of 
$c_i, s_i$ terms. The measured values of $(x_{\pm}, y_{\pm})$ with their 
likelihood contours are shown in Fig.~\ref{fig:bdk_xy}. 

\section{Systematic errors}

Systematic errors in the $x,y$ fit are obtained for the default 
procedure of the combined fit with the optimal binning. 
The systematic errors are summarized in Table~\ref{tab:syst}. 

The uncertainty of the signal shape used in the fit includes the 
following sources: 
\begin{itemize}
  \item Choice of parametrization used to model the shape. 
        The corresponding uncertainty is estimated by using 
        the non-parametric (Keys) function instead of parametrized distribution. 
  \item Possible correlation between the $(\mbc, \de)$ 
        and $(\thr, \fish)$ distributions. To estimate its effect 
        we use 4D binned histogram to describe the distribution. 
  \item MC description of the (\thr, \fish) distribution. Its 
        effect is estimated by floating the parameters 
        of the distribution in the fit to \bdpi control sample. 
  \item Dependence of the signal width on the Dalitz plot bin. 
        The uncertainty due to this effect is estimated 
        by performing the \bdpi fit 
        with the shape parameters floated separately for each bin, and then 
        using the results in the fit to \bdk data. 
\end{itemize}
We do not assign the uncertainty due to the difference in $(\mbc, \de)$ 
shape between the MC and data since the width of the signal distribution 
is calibrated on \bdpi data. 

In the uncertainty of the continuum background shape, the same four 
sources are considered as for the signal distribution. The uncertainty 
due to the choice of parametrization is estimated similarly by using 
the Keys function. The effect of possible correlation between the 
$(\mbc, \de)$ and $(\thr, \fish)$ distributions is estimated
by using the distribution split into the sum of two components 
($u,d,s$ and charm contributions) with independent $(\mbc, \de)$ and 
$(\thr, \fish)$ shapes. The uncertainty due to MC description of 
the $(\mbc, \de)$ and $(\thr, \fish)$ distributions is estimated by 
floating their parameters in the \bdpi fit. To estimate the effect 
of possible correlation of the shape with  the Dalitz plot variables
we fit the shapes separately in each Dalitz plot bin. 

The uncertainties of the shapes of random and peaking $B\overline{B}$
backgrounds are estimated conservatively by performing the fit with 
$\de>-0.1$ GeV --- this requirement rejects the peaking $B\overline{B}$ 
background and a large part of the random $B\overline{B}$ background. 

In the case of the fit of \bdk sample, the uncertainty of the 
\bdpi background shape in $(\thr, \fish)$ variables is estimated 
by taking the $(\thr, \fish)$ shape for signal events. 
The Dalitz plot uncertainty is estimated by using the 
number of flavor-tagged events in bins (rather than the number of \bdpi
events used in the default fit). Uncertainties due to possible 
correlations are treated as in the case of the signal distribution. 

The uncertainty due to Dalitz plot efficiency shape appears because of 
a difference in average efficiency for the flavor and \bdk samples. 
The maximum difference of 1.5\% is obtained in the MC study. 
The uncertainty is obtained from the maximum of two quantities: 
\begin{itemize}
  \item RMS of $x$ and $y$ from smearing the numbers of 
events in the flavor sample $K_i$ by 1.5\%.
  \item Bias of $x$ and $y$ between the fits with and without 
efficiency correction for $K_i$ obtained from signal MC. 
\end{itemize}

The uncertainty due to cross-feed of events between bins is 
conservatively estimated by taking the bias between the fits 
with and without cross-feed. 

The uncertainty arising from the finite sample of flavor-tagged 
$\dkpp$ decays is evaluated by varying the numbers of flavor-tagged 
events in bins $K_i$ within their statistical errors. 

The final results for $x,y$ are corrected for the fit bias obtained from 
the fits of MC pseudo-experiments. The uncertainty due to the fit bias 
is taken from the difference of biases for various input values of $x$ and $y$. 

The uncertainty due to errors of $c_i$ and $s_i$ parameters 
is obtained by smearing the $c_i$ and $s_i$ values within their 
total errors and repeating the fits for the same experimental data. 
We have performed a study of this procedure using the MC pseudo-experiments and 
analytical calculations. We find that the uncertainty obtained 
this way is sample-dependent for small $B$ data samples and its average 
scales inversely proportional to the square root of sample size. 
It reaches a constant value for large $B$ data samples 
(in the systematics-dominated case). This explains a somewhat higher 
uncertainty compared to the CLEO estimate given in~\cite{cleo_2} obtained 
in the limit of very large $B$ sample. 
In addition, the uncertainty in $x,y$ is proportional to $r_B$, and thus 
the uncertainty of the phases $\phi_3$ and $\delta_B$ is independent of 
$r_B$. As a result, the uncertainty of $x,y$ in the \bdk sample fit is 
3--4 times larger than for \bdpi. 

\begin{table*}
  \caption{Systematic errors of $x,y$ measurement for \bdpi and \bdk samples
           in units of $10^{-3}$.}
  \label{tab:syst}
  \begin{tabular}{|l|cccc|cccc|}
    \hline
                                         & \multicolumn{4}{c|}{\bdpi} & \multicolumn{4}{c|}{\bdk} \\

    \cline{2-9}

    Source of uncertainty                & $\Delta x_-$ & $\Delta y_-$ & $\Delta x_+$ & $\Delta y_+$
                                         & $\Delta x_-$ & $\Delta y_-$ & $\Delta x_+$ & $\Delta y_+$ \\
    \hline

    Signal shape                         & $0.9$ & $1.9$  & $1.1$  & $5.0$
                                         & $7.3$ & $7.4$  & $7.3$  & $5.1$ \\
 
    $u,d,s,c$ continuum background       & $0.9$ & $1.4$  & $0.8$  & $1.3$
                                         & $6.7$ & $5.6$  & $6.6$  & $3.2$ \\
 
    $B\overline{B}$ background           & $3.3$ & $1.6$  & $4.5$  & $1.1$
                                         & $7.8$ & $12.2$ & $7.2$  & $6.1$ \\

    \bdpi background                     & $-$   & $-$    & $-$    & $-$
                                         & $1.2$ & $4.2$  & $1.9$  & $1.9$ \\

    Dalitz plot efficiency               & $3.0$ & $1.9$  & $3.2$  & $1.6$
                                         & $4.8$ & $2.0$  & $5.6$  & $2.1$ \\

    Cross-feed between bins              & $0.4$ & $3.0$  & $0.7$  & $0.9$
                                         & $0.4$ & $9.0$  & $0.6$  & $3.0$ \\

    Flavor-tagged statistics             & $1.7$ & $2.0$  & $1.6$  & $2.0$
                                         & $1.5$ & $2.7$  & $1.7$  & $1.9$ \\

    Fit bias                             & $0.4$ & $0.5$  & $0.4$  & $0.5$
                                         & $3.2$ & $5.8$  & $3.2$  & $5.8$ \\

    $c_i, s_i$ precision                 & $2.6$ & $6.5$  & $2.6$  & $6.5$
                                         & $10.1$& $22.5$ & $7.2$  & $17.4$\\

    \hline

    Total without $c_i$,$s_i$ precision  & $ 5.0$& $5.0$  & $6.0$  &$6.0$
                                         & $14.0$& $19.4$ & $14.0$ &$11.3$ \\

    \hline

    Total                                & $5.6$ & $8.2$  & $ 6.5$ & $8.8$
                                         & $17.3$& $29.7$ & $15.7$ & $20.7$ \\

    \hline
  \end{tabular}
\end{table*}

\section{Results for $\phi_3$, $r_B$ and $\delta_B$}

We use frequentist treatment with the Feldman-Cousins ordering to 
obtain the physical parameters $\mu = (\phi_3, r_B, \delta_B)$ from the 
measured parameters $z=(x_-, y_-, x_+, y_+)$ as was done in the 
previous Belle analyses. In essence, the confidence level $\alpha$
for a set of physical parameters $\mu$ is calculated as 
\begin{equation}
  \alpha(\mu) = \int\limits_{\mathcal{D}(\mu)}p(z|\mu)dz\left/ 
                \int\limits_{\infty}p(z|\mu)dz\right., 
\end{equation}
where $p(z|\mu)$ is the probability density to obtain the measurement 
result $z$ given the set of physics parameters $\mu$. The integration 
domain $\mathcal{D}(\mu)$ is given by the likelihood ratio (Feldman-Cousins)
ordering: 
\begin{equation}
  \frac{p(z|\mu)}{p(z|\mu_{\rm best}(z))} > 
  \frac{p(z_0|\mu)}{p(z_0|\mu_{\rm best}(z_0))}, 
\end{equation}
where $\mu_{\rm best}(z)$ is $\mu$ that maximizes $p(z|\mu)$ 
for the given $z$, and $z_0$ is the result of the data fit. 

The difference with the previous Belle analyses is that the probability 
density $p(z|\mu)$ is a multivariate Gaussian PDF with the errors 
and correlations between $x_{\pm}$ and $y_{\pm}$ taken from the 
data fit result. In the previous analyses, this PDF was taken from 
MC pseudo-experiments. 

\begin{figure}
  \parbox[t]{0.5\textwidth}{
  \includegraphics[width=0.242\textwidth]{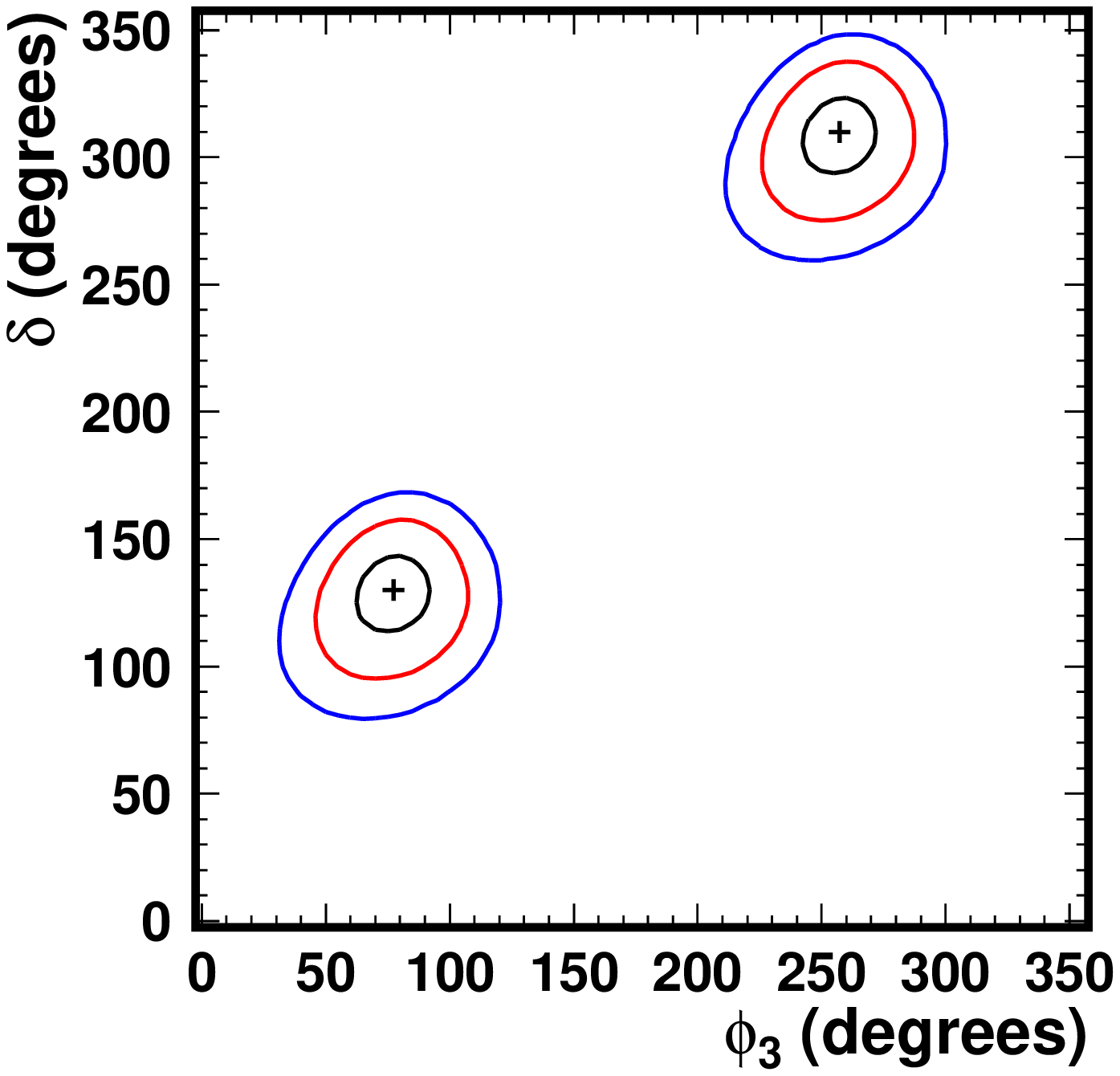}
  \includegraphics[width=0.242\textwidth]{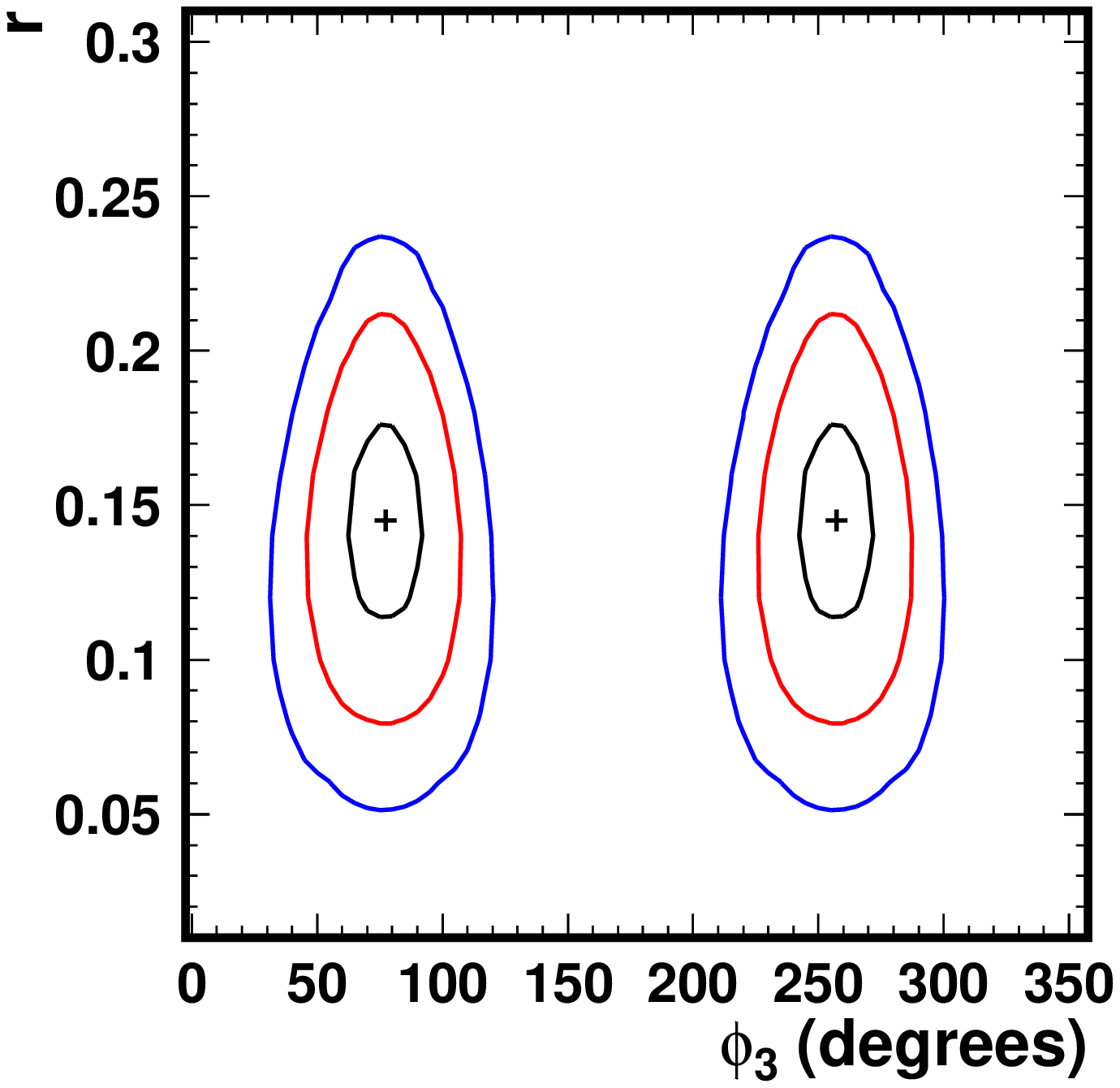}
  \put(-225,102){(a)}
  \put(-97,102){(b)}
  }
  \caption{Two-dimensional projections of 
           confidence region onto $(\phi_3, \delta_B)$ and 
           $(\phi_3, r_B)$ planes (one-, two-, and three standard 
           deviations). }
  \label{fig:bdk_cl}
\end{figure}

As a result of this procedure, we obtain the confidence levels (CL) 
for the set of 
physical parameters $\phi_3, r_B, \delta_B$. The confidence levels 
for one and two standard deviations are taken at 20\% and 74\%
(the case of three-dimensional Gaussian distribution). The projections 
of the 3D surfaces bounding one and two standard deviations 
volumes onto $\phi_3$ variable, and $(\phi_3, r_B)$ and $(\phi_3, \delta_B)$
planes are shown in Fig.~\ref{fig:bdk_cl}. 

Systematic errors in $\mu$ are obtained by varying the measured 
parameters $z$ within their systematic errors (Gaussian distribution is taken) 
and calculating the RMS of $\mu_{\rm best}(z)$. In this calculation we 
assume that the systematic errors are uncorrelated. In the case of 
$c_i, s_i$ systematics, we test that assumption: when the fluctuation 
in $c_i$ and $s_i$ is generated, we perform the fits to both $B^+$ and 
$B^-$ data with the same fluctuated $c_i, s_i$. We observe no 
significant correlation between resulting $x_-$ and $x_+$ ($y_-$ and $y_+$). 

The final results are:
\begin{equation}
\begin{split}
\phi_3 & = (77.3^{+15.1}_{-14.9} \pm 4.2 \pm 4.3)^{\circ}\\
r_B & = 0.145\pm 0.030 \pm 0.011\pm 0.011\\
\delta_B & = (129.9\pm 15.0\pm 3.9\pm 4.7)^{\circ}, 
\end{split}
\end{equation}
where the first error is statistical, the second is systematic 
error without $c_i, s_i$ uncertainty, 
and the third error is due to $c_i,s_i$ uncertainty.

We do not calculate the statistical significance of $CP$ violation as 
it is done in the previous analyses by taking the CL for $\phi_3=0$: 
this number is purely based on the behavior of the tails of $p(z|\mu)$ 
distribution far from the central value, and Gaussian assumption can lead 
to overestimation of $CP$ violation significance. As a preliminary 
number we use the estimate of probability of the fluctuation 
in the difference of number of events in bins for $B^+$ and $B^-$
data: the probability of such fluctuation in the case of 
$CP$ conservation is $p=0.42$\%. 

\section{Conclusion}

We report the results of a measurement of the unitarity triangle angle 
$\phi_3$ using a model-independent Dalitz plot analysis of
\dkpp\ decay in the process \bdk. 
The measurement was performed 
with a full data sample of 711 fb$^{-1}$ 
($772\times 10^6$ $B\overline{B}$ pairs) collected by the Belle detector
at $\Upsilon(4S)$. 
The model independence is reached by binning the Dalitz plot of \dkpp 
decay and using the strong phase coefficients for bins measured by 
CLEO experiment~\cite{cleo_2}. 
We obtain the value
$\phi_3 = (77.3^{+15.1}_{-14.9} \pm 4.2 \pm 4.3)^{\circ}$; 
of the two possible solutions we choose the one with $0<\phi_3<180^{\circ}$.
We also obtain the value of the amplitude ratio 
$r_B = 0.145\pm 0.030 \pm 0.011\pm 0.011$. 
These results are preliminary. 

This analysis is a first application of the novel method of $\phi_3$
measurement. Although currently it does not offer significant 
advantages over the model-dependent Dalitz plot analyses of the same 
decay chain, it is promising for the measurement at super-B 
factories~\cite{superkekb, superb}. 
We expect that the statistical error of the $\phi_3$ measurement 
using the statistics of 50 ab$^{-1}$ to be available at the super-B factory 
will reach $1-2^{\circ}$. With the use of BES-III data~\cite{besiii}
the error due to the phase terms of \dkpp decay will decrease to 
$1^{\circ}$ or less. We also expect that the experimental systematic 
error can be kept at the level below $1^{\circ}$ since most of 
its sources are limited by the statistics of the control channels.

\section*{Acknowledgments}

We thank the KEKB group for the excellent operation of the
accelerator, the KEK cryogenics group for the efficient
operation of the solenoid, and the KEK computer group and
the National Institute of Informatics for valuable computing
and SINET4 network support.  We acknowledge support from
the Ministry of Education, Culture, Sports, Science, and
Technology (MEXT) of Japan, the Japan Society for the 
Promotion of Science (JSPS), and the Tau-Lepton Physics 
Research Center of Nagoya University; 
the Australian Research Council and the Australian 
Department of Industry, Innovation, Science and Research;
the National Natural Science Foundation of China under
contract No.~10575109, 10775142, 10875115 and 10825524; 
the Ministry of Education, Youth and Sports of the Czech 
Republic under contract No.~LA10033 and MSM0021620859;
the Department of Science and Technology of India; 
the BK21 and WCU program of the Ministry Education Science and
Technology, National Research Foundation of Korea,
and NSDC of the Korea Institute of Science and Technology Information;
the Polish Ministry of Science and Higher Education;
the Ministry of Education and Science of the Russian
Federation and the Russian Federal Agency for Atomic Energy;
the Slovenian Research Agency;  the Swiss
National Science Foundation; the National Science Council
and the Ministry of Education of Taiwan; and the U.S.\
Department of Energy.
This work is supported by a Grant-in-Aid from MEXT for 
Science Research in a Priority Area (``New Development of 
Flavor Physics''), and from JSPS for Creative Scientific 
Research (``Evolution of Tau-lepton Physics'').

This research is partially funded by the Russian Presidential Grant 
for support of young scientists, grant number MK-1403.2011.2.

\end{document}